\documentclass[preprintnumbers,showpacs,amsmath,amssymb,floatfix,9pt,prd,onecolumn,
superscriptaddress,nofootinbib]{revtex4}
\usepackage{latexsym}
\usepackage{epsfig}
\usepackage{amssymb}
\begin{document}

\title{  Particles and scalar waves  in noncommutative charged black hole spacetime}

\begin{abstract}
In this paper we have discussed  geodesics  and the motion of test particle in the gravitational field of noncommutative charged black hole spacetime.   The motion of massive and massless particle have been discussed seperately. A comparative study of noncommutative charged black hole and usual Reissner-Nordstr\"{o}m black hole has been done. The study of effective potential has also been included. Finally, we have examined the scattering of scalar waves in noncommutative charged black hole spacetime.
\end{abstract}

\author{{  Piyali Bhar}}
\email{piyalibhar90@gmail.com  } \affiliation{ {  Department of
Mathematics, Jadavpur University, Kolkata 700 032, West Bengal,
India}}

 \author{{ Farook Rahaman}}
\email{rahaman@iucaa.ernet.in} \affiliation{ {  Department of
Mathematics, Jadavpur University, Kolkata 700 032, West Bengal,
India}}

\author{Ritabrata Biswas}
\email{biswas.ritabrata@gmail.com  } \affiliation{Department of
Mathematics, Krishnagar Women's College, Arobinda Sarani,
Krishnanagar, Nadia - 741101, West Bengal, India }

\author{U. F. Mondal}
\email{umarfarooquemondal@ymail.com} \affiliation{ Department of
Mathematics, Behala College, Parnasree, Kolkata 700060, India }

\date{\today}

\maketitle

\section{Introduction}
Noncommutative geometry (NCG) is concerned with a geometric
approach to noncommutative algebras, and with the construction of
spaces that are locally presented by noncommutative algebras of
functions. Use of such a geometry in cosmological purpose came
into existence after the works of Chamseddine,
\cite{Chamseddine1,kl}. A general D-Dimensional static spherically
symmetric solutions of the specific vector model coupled to
Einstein gravity was proposed by \cite{Klimcik1}. Impact of non
commutative geometry on different black holes have been studied
by different authors \cite{Nasseri1}.  Surprisingly, it was shown
that thermodynamic behavior of the noncommutative Schwarzschild
black hole is analogous to that of the Reissner-Nordstrom ( RN )
black hole   in the near extremal limit \cite{Kim1}. Alavi worked
on a radiating RN black hole in non commutative geometry and
establish the facts : the
 existence of a minimal non-zero mass to which the black hole can
shrink; a
finite maximum temperature that the black hole can reach before cooling down
to absolute zero; compared to the neutral black holes the effect of charge is to
increase the minimal non-zero mass and lower the maximum temperature;
the absence of any curvature singularity \cite{Alavi1}. Three Dimensional Charged Black Hole Inspired by Noncommutative Geometry was studied in \cite{Larranaga1}. Charged non rotating black holes are been studied in \cite{Modesto1}. In \cite{Larranaga2} by considering particles as smeared objects, Larranaga investigate the effects of space noncommutativity on the geodesic structure in Schwarzschild-AdS spacetime. Considering the effects of noncommutativity in the orbits of particles in
Schwarzschild-AdS spacetime it is found that there do exist some For radial time-like geodesics, there are some bounded trajectories. For non-radial time-like geodesics, elliptical orbits are allowed as well as circular orbits.

The usual semiclassical laws of physics break down at plank scale and at this circumstances noncommutativity comes into consideration. The noncommutativity of the spacetime can be encoded by the commutator $[x^{\mu},x^{\nu}]=i\theta^{\mu \nu}$ where $\theta^{\mu \nu}$ is the anti-symmetric metric.In literature \cite{smailagic}\cite{Rizzo} it has been shown that noncommutativity replaces the point-like structure to a smeared objects. The usual definition of mass density function of the commutative space namely Dirac delta function is not so far valid in noncommutative spacetime. In noncommutative space the Dirac delta function is replaced by the gaussian distribution function of the form \cite{Nicolini}
\begin{equation}
\rho_{\theta}=\frac{M}{(4\pi \theta)^{\frac{3}{2}}}e^{-\frac{r^{2}}{4\theta}}
\end{equation}
where $\theta$ is the noncommutative parameter and is of dimension $length^{2}$.\\

 ~~~~~Several authors  have described different astrophysical phenomena in noncommutative spacetime. Regular black hole in $3D$ noncommutative spacetime is analyzed by Myung et al \cite{Myung}.  Baberjee et al \cite{banerjee} have studied noncommutative Schwarzschild Black hole and area law. Three dimensional charged black hole inspired by noncommutative geometry has been discussed in \cite{Alexis}. Rahaman et al  \cite{rahaman} studied BTZ black hole inspired by noncommutative geometry.
Galactic rotation curve in noncommutative geometry background has been analyzed in \cite{farook1}. Kuhfittig\cite{Kuhfittig} found that a special class of thin shell wormhole that are unstable in classical general relativity but they in a small region in noncommutative spacetime. Higher dimensional wormhole in noncommutative geometry has been discussed in \cite{farook2}. In that paper authors have shown that the wormhole exists only in four and five dimensional spacetime. ~Inspired by some of the earlier works  \cite{Nicolini}~\cite{grezia}~\cite{chabab},  in this paper,  we will consider noncommutative charged black hole \cite{Ansoldi}. The underlying fluid is anisotropic in nature but in the limit $r>>\sqrt{\theta}$ and $r<<\sqrt{\theta}$ it becomes commutative. We shall compare Geodesic study of this black hole with the  usual Reissner-Nordstr\"{o}m black hole. Next,  we will discuss about the motion of a test particle in  noncommutative charged black hole spacetime. Finally, we  examine  the scattering of scalar waves in noncommutative charged black hole spacetime.  \\

 ~~~~The plan of the present paper as follows: In section II we have discussed about noncommutative charged black hole black hole. In section III geodesic study has been done (massless and massive particle have been discussed separately ).  In section IV,  we have discussed about effective potential and in section V motion of test particle have been studied. In section VI,  we have   examined  the scattering of scalar waves in noncommutative charged black hole spacetime and lastly some concluding remarks have been included.

\section{ Noncommutative charged black hole}
The metric of a noncommutative charged black hole is described by the metric given by \cite{Ansoldi},
\begin{equation}
ds^{2}=-\left(1-\frac{2M_{\theta}}{r}+\frac{Q_{\theta}^{2}}{r^{2}}\right)dt^{2}+\left(1-\frac{2M_{\theta}}{r}
+\frac{Q_{\theta}^{2}}{r^{2}}\right)^{-1}dr^{2}+r^{2}d\Omega^{2}
\end{equation}
where
\[M_{\theta}(r)=\frac{2M}{\sqrt{\pi}}\gamma\left(\frac{3}{2},\frac{r^{2}}{4\theta}\right)\]
\[Q_{\theta}(r)=\frac{Q}{\sqrt{\pi}}\sqrt{\gamma^{2}\left(\frac{1}{2},\frac{r^{2}}{4\theta}\right)
-\frac{r}{\sqrt{2\theta}}\gamma\left(\frac{1}{2},\frac{r^{2}}{2\theta}\right)}\]
\begin{equation}
\gamma\left(\frac{a}{b},x\right)=\int_0^{x}u^{\frac{a}{b}-1}e^{-u}du
\end{equation}
The horizon radius is $r_h$ where $f(r_h)=0$ i.e.  where $f(r)$ cuts the $r$ axis  (see $ fig.1 $).

Here,
\begin{equation}
f(r)=1-\frac{4M}{r\sqrt{\pi}}\gamma\left(\frac{3}{2},\frac{r^{2}}{4\theta}\right)
+\frac{Q^{2}}{r^{2}\pi}\left[\gamma^{2}\left(\frac{1}{2},\frac{r^{2}}{4\theta}\right)
-\frac{r}{\sqrt{2\theta}}\gamma\left(\frac{1}{2},\frac{r^{2}}{2\theta}\right)\right]
\end{equation}

where $M$ is the mass and $Q$ is the charge of the   black hole.
It is noted that  for large r, Reissner-Nordstr\"{o}m black hole will be obtained.

The horizon radius    ( $r_h $)  can be found where $f(r_h) = 0$ in other words, where   f(r)
cuts r axis (see fig-1). We observe that for specific values of the parameters, say, M=2, Q=1 and $\theta=0.1$, we get two raii of horizons where outer  horizon   exists at $r_h= 0.9267$.

\begin{figure}[htbp]
    \centering
        \includegraphics[scale=.3]{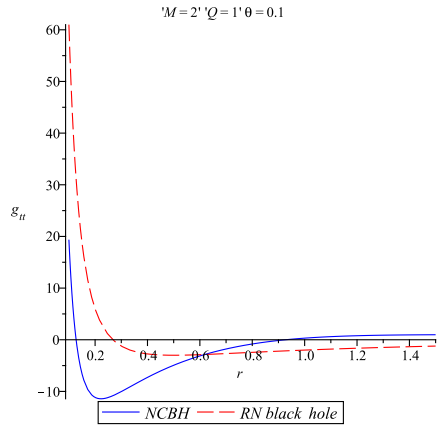}
       \caption{Two horizons exist for  Noncommutative charged black hole. }
    \label{fig:3}
\end{figure}

\section{The Geodesics}
Without loss of generality,  let us consider the geodesic motion in the plane $\vartheta=\frac{\pi}{2}$.\\
From the geodesics equation
\begin{equation}
\frac{d{x^{\mu}}}{d\tau^{2}}+\Gamma_{\nu \lambda}^{\nu}\frac{dx^{\nu}}{d\tau}\frac{dx^{\lambda}}{d\tau}=0,
\end{equation}
one can obtain
\begin{equation}
\frac{1}{f(r)}\left(\frac{dr}{d\tau}\right)^{2}=\frac{E^{2}}{f(r)}-\frac{J^{2}}{r^{2}}-L
\end{equation}
\begin{equation}
r^{2}\left(\frac{d\phi}{d\tau}\right)=J
\end{equation}
and
\begin{equation}
\frac{dt}{d\tau}=\frac{E}{f(r)}
\end{equation}
where f(r) is given in (4)  and the constants $E~and~J $ are energy per unit mass and angular momentum respectively about an axis invariant to the plane $\theta=\frac{\pi}{2}$. Here $\tau$ is the affine parameter and $L $ is the Lagrangian having value $0~and~1$ respectively for massless and massive particle.\\

Now for radial geodesic $(J=0)$.
Then from equation $(6)$ we get,
\begin{equation}
\left(\frac{dr}{d\tau}\right)^{2}=E^{2}-L f(r)
\end{equation}
Using the   equations (8) and (9),  we get,
\begin{equation}
\left(\frac{dr}{dt}\right)^{2}=\left\{f(r)\right\}^{2}\left[1-\frac{L}{E^{2}}f(r)\right]
\end{equation}
\subsection{For photon Like Particle (L=0)}
For photon like particle we have,
\begin{equation}
\left(\frac{dr}{dt}\right)^{2}=\left\{1-\frac{4M}{r\sqrt{\pi}}\gamma\left(\frac{3}{2},\frac{r^{2}}{4\theta}\right)
+\frac{Q^{2}}{r^{2}\pi}\left[\gamma^{2}\left(\frac{1}{2},\frac{r^{2}}{4\theta}\right)
-\frac{r}{\sqrt{2\theta}}\gamma\left(\frac{1}{2},\frac{r^{2}}{2\theta}\right)\right]\right\}^{2},
\end{equation}
which gives,
\begin{equation}
\pm t=\int_{r_h}^{r^{*}}\frac{dr}{1-\frac{4M}{r\sqrt{\pi}}\gamma\left(\frac{3}{2},\frac{r^{2}}{4\theta}\right)
+\frac{Q^{2}}{r^{2}\pi}\left[\gamma^{2}\left(\frac{1}{2},\frac{r^{2}}{4\theta}\right)
-\frac{r}{\sqrt{2\theta}}\gamma\left(\frac{1}{2},\frac{r^{2}}{2\theta}\right)\right]}
\end{equation}
The above integration can not be performed analytically but  by the help of numerical integration one can obtain the values of $t$ for different values of $r $ which has shown in TABLE-$1$.

\begin{table}
\caption{Values of $t$  for different $r $. $ r_h =0.9267 $,
$\theta=0.1$,$M=2$,$Q=1$ }

{\begin{tabular}{@{}cc@{}} \toprule $r $ & $t$ \\
\colrule
2   &  ~~~~1.2014\\
2.5 &  ~~~~ 1.7014\\
3   &  ~~~~2.2014\\
3.5 &  ~~~~2 .7014\\
4   &  ~~~~3.2014\\
4.5   &  ~~~~3.7014\\
 \botrule
\end{tabular}}
\end{table}

The relation between the distance$(r)$ and time $(t)$ has been plotted in $\textbf{fig.2}$.

From equation $(9)$ we get,
\begin{equation}
\left(\frac{dr}{d\tau}\right)^{2}=E^{2}
\end{equation}
which gives,
\begin{equation}
r=\pm E\tau
\end{equation}
The relation between the distance $r$ and proper time $(\tau)$ has been plotted in $\textbf{fig.2}$

\subsection{Massive Particle Motion(L=1)}

From equations (6) and (8), we obtain
\begin{equation}
\left(\frac{dr}{dt}\right)^{2}=\left(\frac{f(r)}{E^{2}}\right)^{2}(E^{2}-f(r))
\end{equation}
which implies,
\begin{equation}
\pm t=\int_{r_h}^{r^{*}}\frac{E dr}{\left\{1-\frac{4M}{r\sqrt{\pi}}\gamma\left(\frac{3}{2},\frac{r^{2}}{4\theta}\right)
+\frac{Q^{2}}{r^{2}\pi}\left[\gamma^{2}\left(\frac{1}{2},\frac{r^{2}}{4\theta}\right)
-\frac{r}{\sqrt{2\theta}}\gamma\left(\frac{1}{2},\frac{r^{2}}{2\theta}\right)\right]\right\}
\sqrt{E^{2}-1+\frac{4M}{r\sqrt{\pi}}\gamma\left(\frac{3}{2},\frac{r^{2}}{4\theta}\right)
-\frac{Q^{2}}{r^{2}\pi}\left[\gamma^{2}\left(\frac{1}{2},\frac{r^{2}}{4\theta}\right)
-\frac{r}{\sqrt{2\theta}}\gamma\left(\frac{1}{2},\frac{r^{2}}{2\theta}\right)\right]}}
\end{equation}
As before, the above integration can not be obtained analytically, so we  obtain the numerical values
of $t$ for different values of $r $ which has shown in TABLE-$2$\\

\begin{table}
\caption{Values of $t$  for different $r $. $ r_h =0.9267 $,
$\theta=0.1$,$M=2$,$Q=1$,$E=1.5$ }

{\begin{tabular}{@{}cc@{}} \toprule $r $ & $t$ \\
\colrule
2   &  ~~~~1.2340\\
2.5 &  ~~~~1.7504\\
3   &  ~~~~2.2668\\
3.5 &  ~~~~2.7832\\
4.5   &  ~~~~3.8160\\
 5 &  ~~~~4.3324\\
\botrule
\end{tabular}}
\end{table}

The relation between the distance$(r)$ and time $(t)$ has been plotted in $\textbf{fig.3}$\\

From equation $(9)$ we get,
\begin{equation}
\left(\frac{dr}{d\tau}\right)^{2}=E^{2}-f(r)
\end{equation}
which gives,
\begin{equation}
\pm \tau=\int_{r_h}^{r }\frac{dr}{\sqrt{E^{2}-1+
\frac{4M}{r\sqrt{\pi}}\gamma\left(\frac{3}{2},\frac{r^{2}}{4\theta}\right)
-\frac{Q^{2}}{r^{2}\pi}\left[\gamma^{2}\left(\frac{1}{2},\frac{r^{2}}{4\theta}\right)
-\frac{r}{\sqrt{2\theta}}\gamma\left(\frac{1}{2},\frac{r^{2}}{2\theta}\right)\right]}}
\end{equation}

\begin{table}
\caption{Values of $t$  for different $r $. $ r_h =0.9267 $,
$\theta=0.1$,$M=2$,$Q=1$,$E=1.5$ }

{\begin{tabular}{@{}cc@{}} \toprule $r $ & $t$ \\
\colrule
2   &   ~~~~0.9108\\
2.5 &  ~~~~ 1.3580 \\
3   &  ~~~~1.8052\\
3.5 &  ~~~~2.2524\\
4  &  ~~~~2.6996\\
4.5 &  ~~~~3.1468\\
\botrule
\end{tabular}}
\end{table}

Again, the  above integration can not be solved analytically, so we   obtain the numerical values of $t$ for different values of $r $ which has shown in TABLE. $3$\\
The relation between the distance$(r)$ and proper time $(\tau)$ has been plotted in $\textbf{fig.3}$.\\

\begin{figure*}[thbp]
\begin{tabular}{rl}
\includegraphics[width=5.0cm]{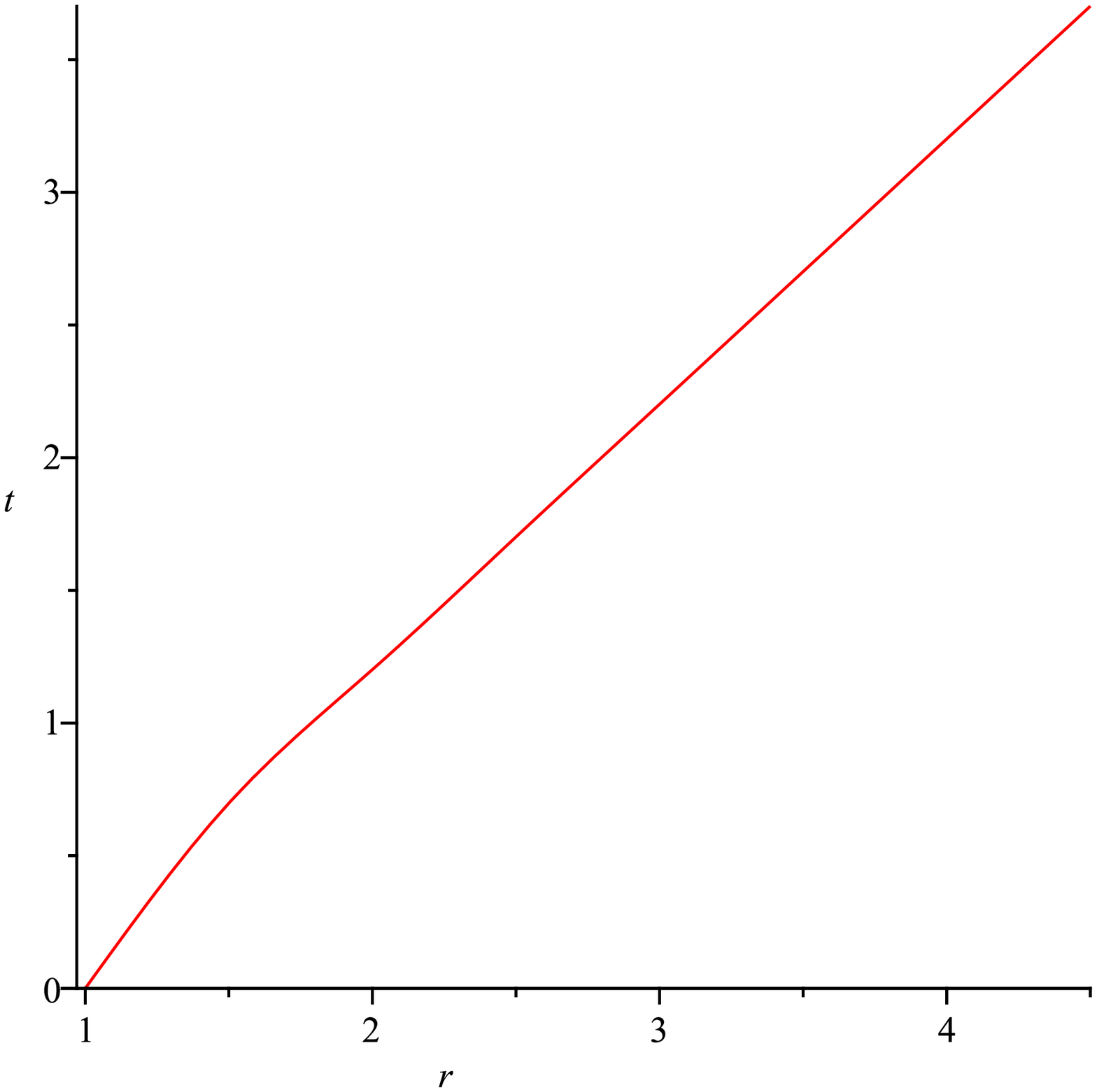}&
\includegraphics[width=5.0cm]{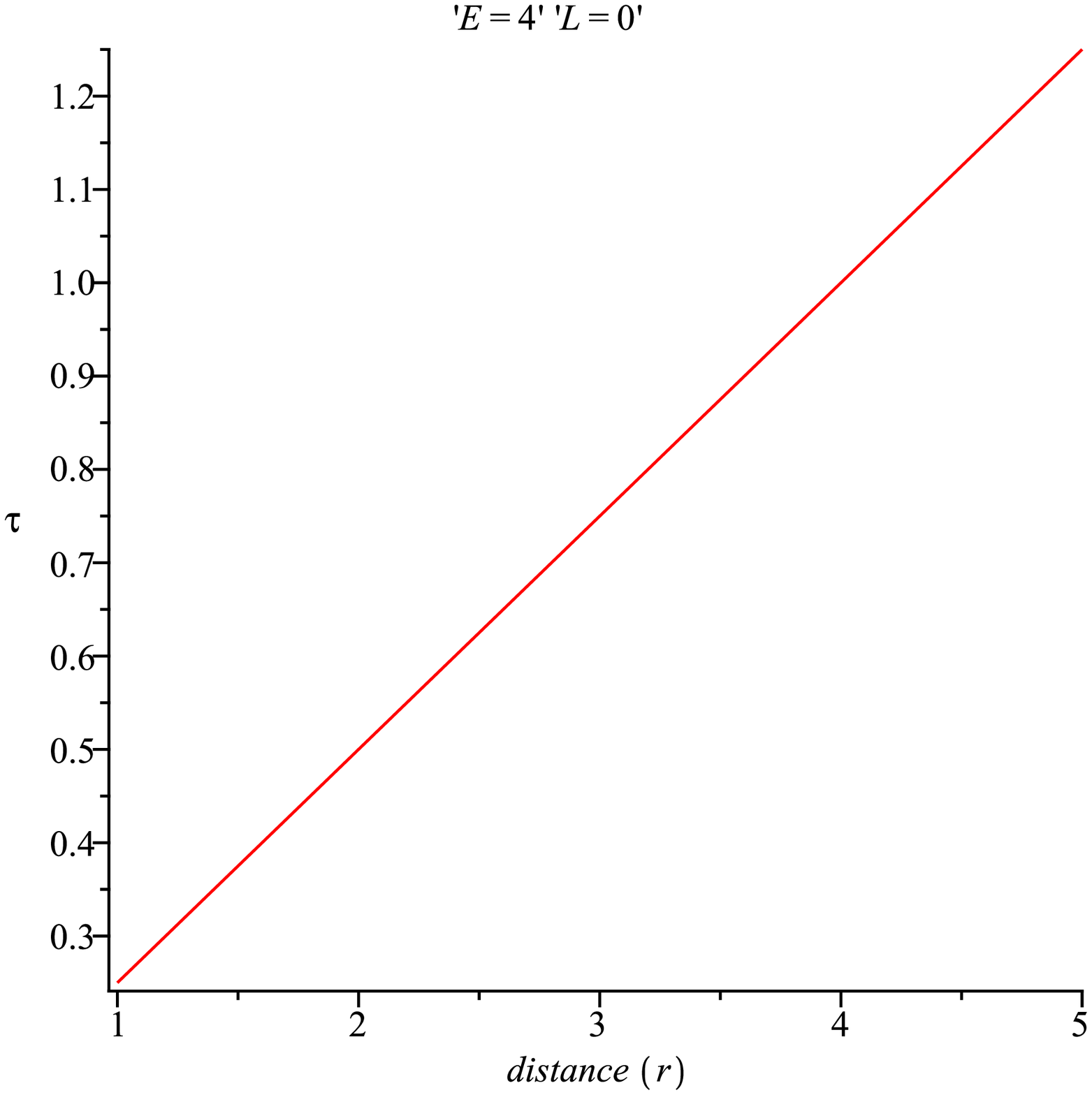}\\
\end{tabular}
\caption{ (Left) Variation of $t$ has been plotted against $r$ for massless particle.
 (Right) Variation of proper time $ (\tau) $  has been plotted against $r$ for massless particle. }
\end{figure*}

\begin{figure*}[thbp]
\begin{tabular}{rl}
\includegraphics[width=5.0cm]{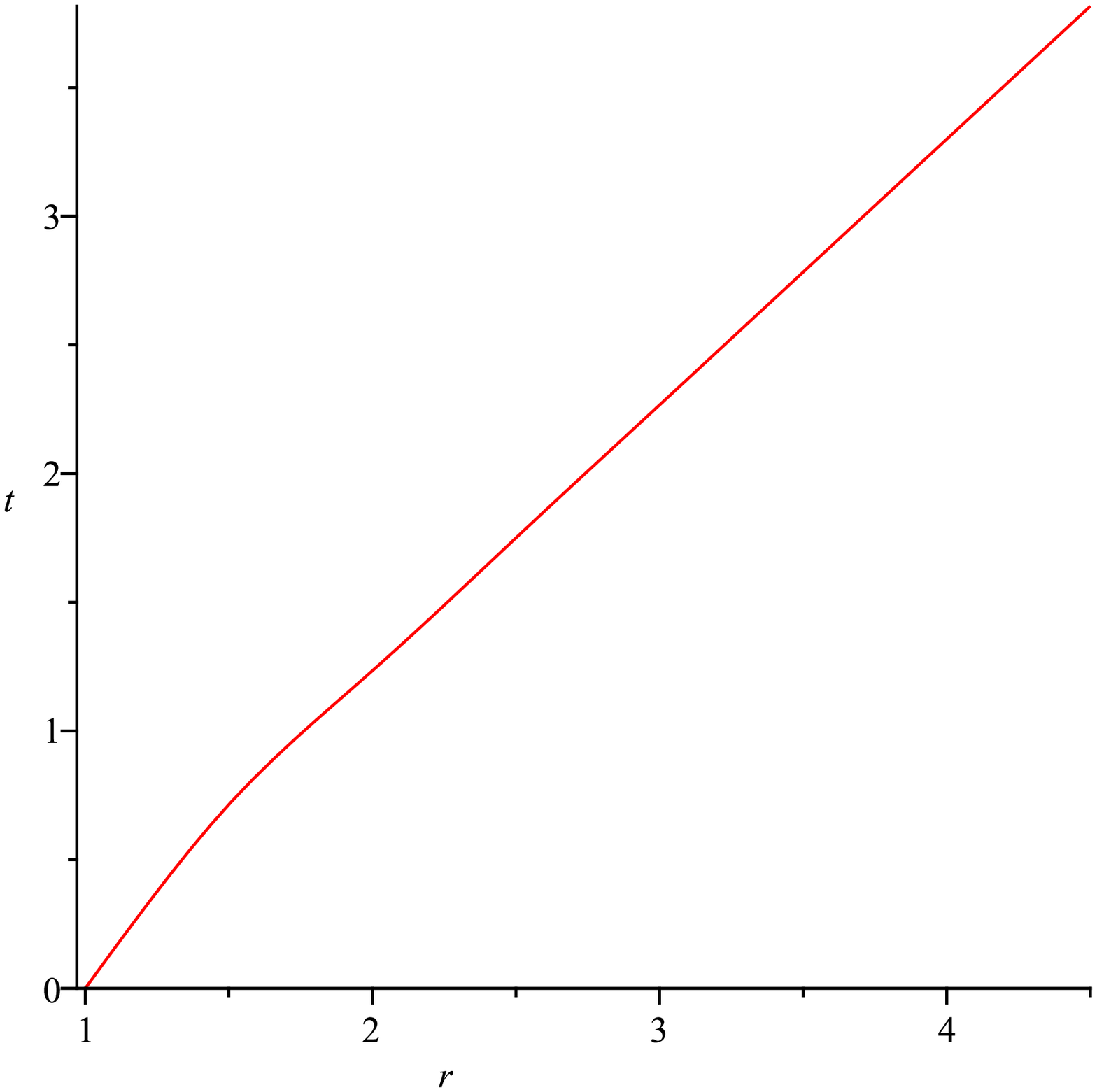}&
\includegraphics[width=5.0cm]{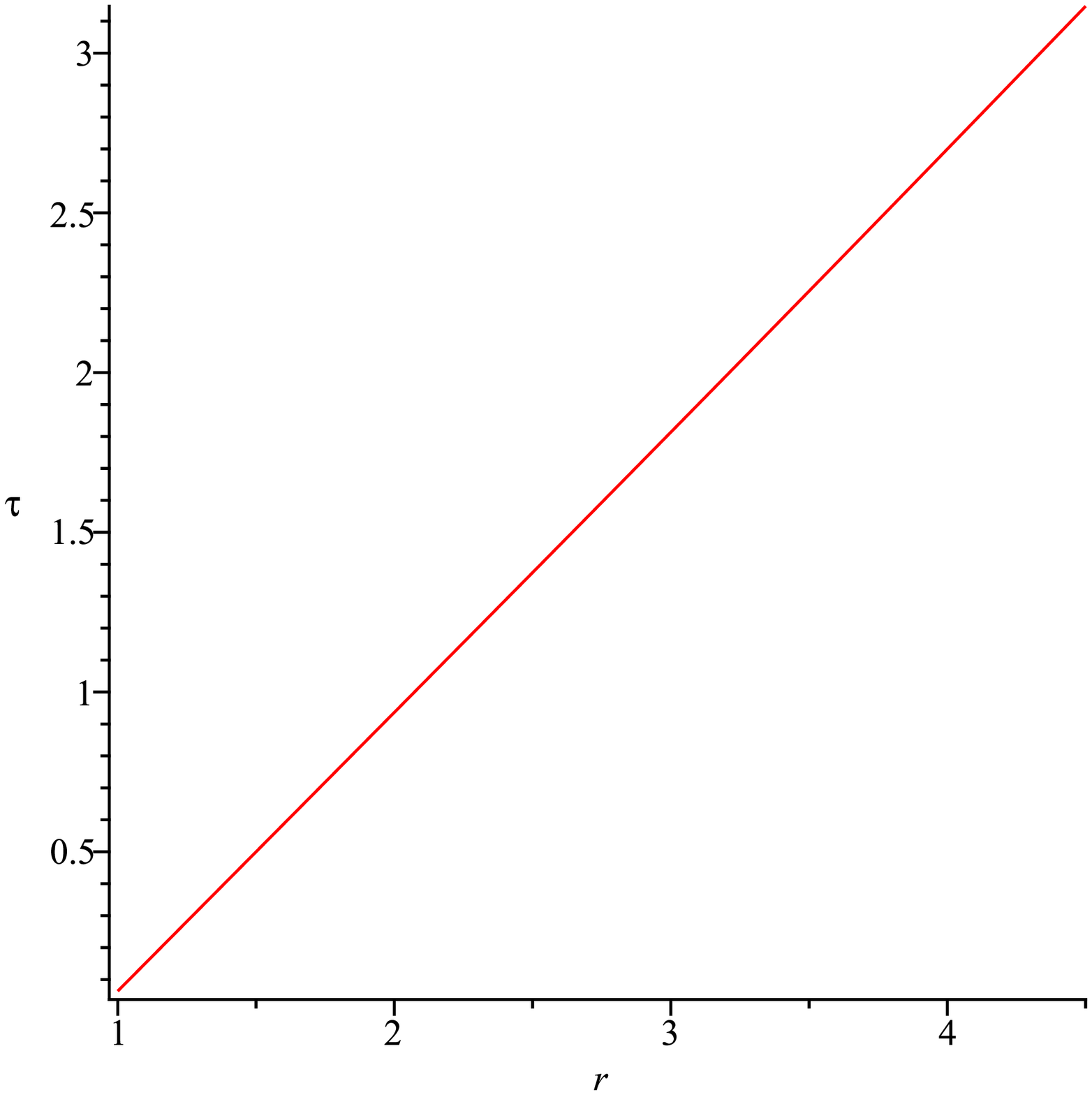}\\
\end{tabular}
\caption{ (Left) Variation of $t$ has been plotted against $r$ for massive particle.
 (Right) Variation of proper time $ (\tau) $  has been plotted against $r$ for massive particle. }
\end{figure*}

\section{Effective Potential}
From equation $(6)$ we get,
\begin{equation}
\frac{1}{2}\left(\frac{dr}{d\tau}\right)^{2}=\frac{1}{2}\left[E^{2}-f(r)\left(L+\frac{J^{2}}{r^{2}}\right)\right]
\end{equation}
Now comparing the equation $()$ with the equation $\frac{\dot{r}^{2}}{2}+V_{eff}=0$,one can get the effective potential as,
\begin{equation}
V_{eff}=-\frac{1}{2}\left[E^{2}-f(r)\left(L+\frac{J^{2}}{r^{2}}\right)\right]
\end{equation}

\subsection{For Photonlike Particle (L=0)}
Now for radial geodesics $p=0$. In that case $V_{eff}$ is given by
\[V_{eff}=-\frac{1}{2}E^{2}\]
Now if $E=0,~V_{eff}$ becomes zero and the particle behaves "like a free" particle.
So we will consider the case
$p \neq 0. $

Then $V_{eff }$ becomes,
\begin{equation}
V_{eff}=-\frac{E^{2}}{2}+\frac{J^{2}}{2r^{2}}\left[1-\frac{4M}{r\sqrt{\pi}}\gamma\left(\frac{3}{2},\frac{r^{2}}{4\theta}\right)
+\frac{Q^{2}}{r^{2}\pi}\left\{\gamma^{2}\left(\frac{1}{2},\frac{r^{2}}{4\theta}\right)
-\frac{r}{\sqrt{2\theta}}\gamma\left(\frac{1}{2},\frac{r^{2}}{2\theta}\right)\right\}\right]
\end{equation}

As $r\rightarrow 0 $ the effective potential $V_{eff}$ becomes very large. As $r\rightarrow\infty,~~V_{eff}$ becomes $-\frac{E^{2}}{2}$ ~~and at the horizon $V_{eff}=-\frac{E^{2}}{2}$\\

\begin{figure*}[thbp]
\begin{tabular}{rl}
\includegraphics[width=5.0cm]{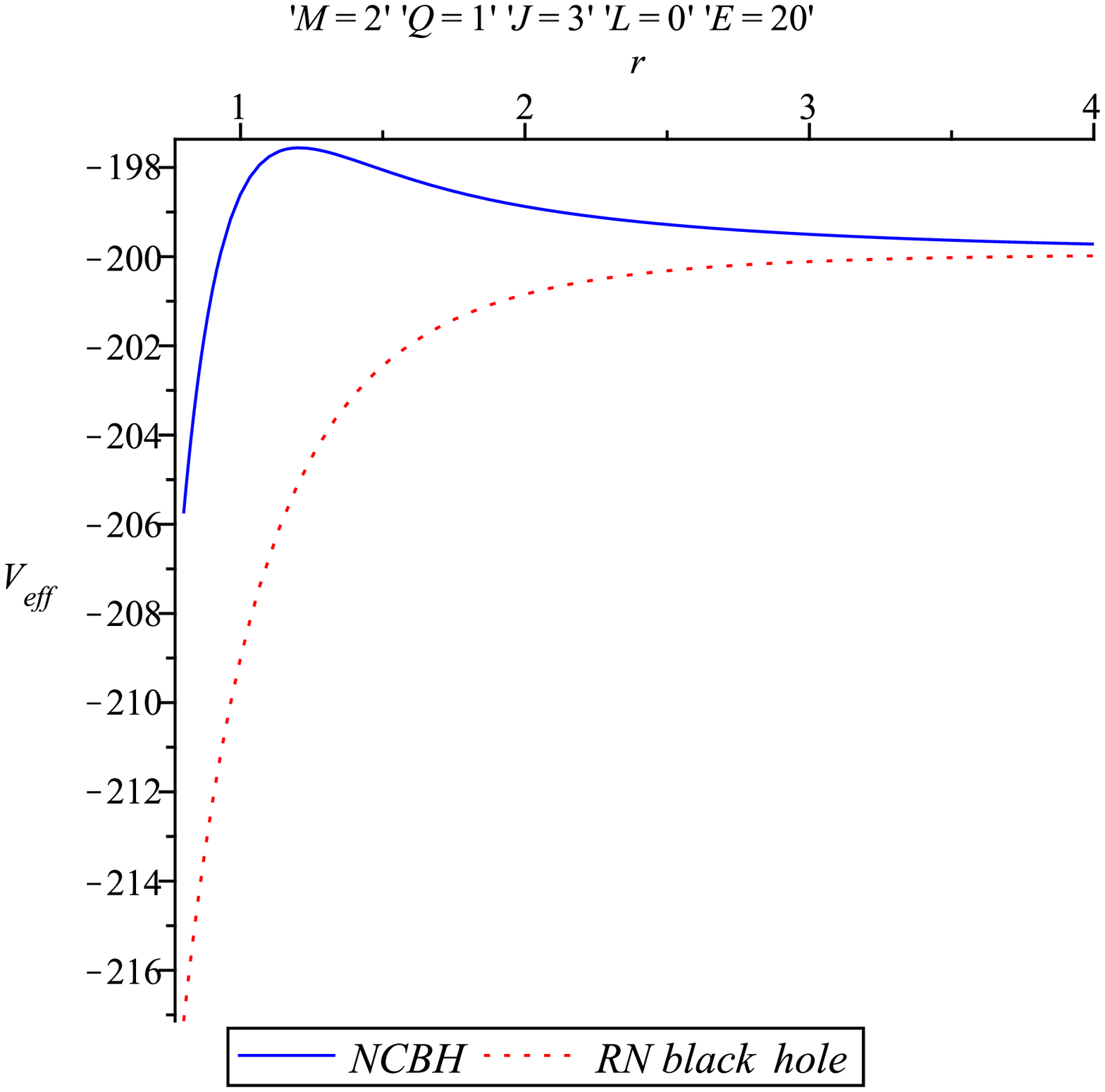}&
\includegraphics[width=5.0cm]{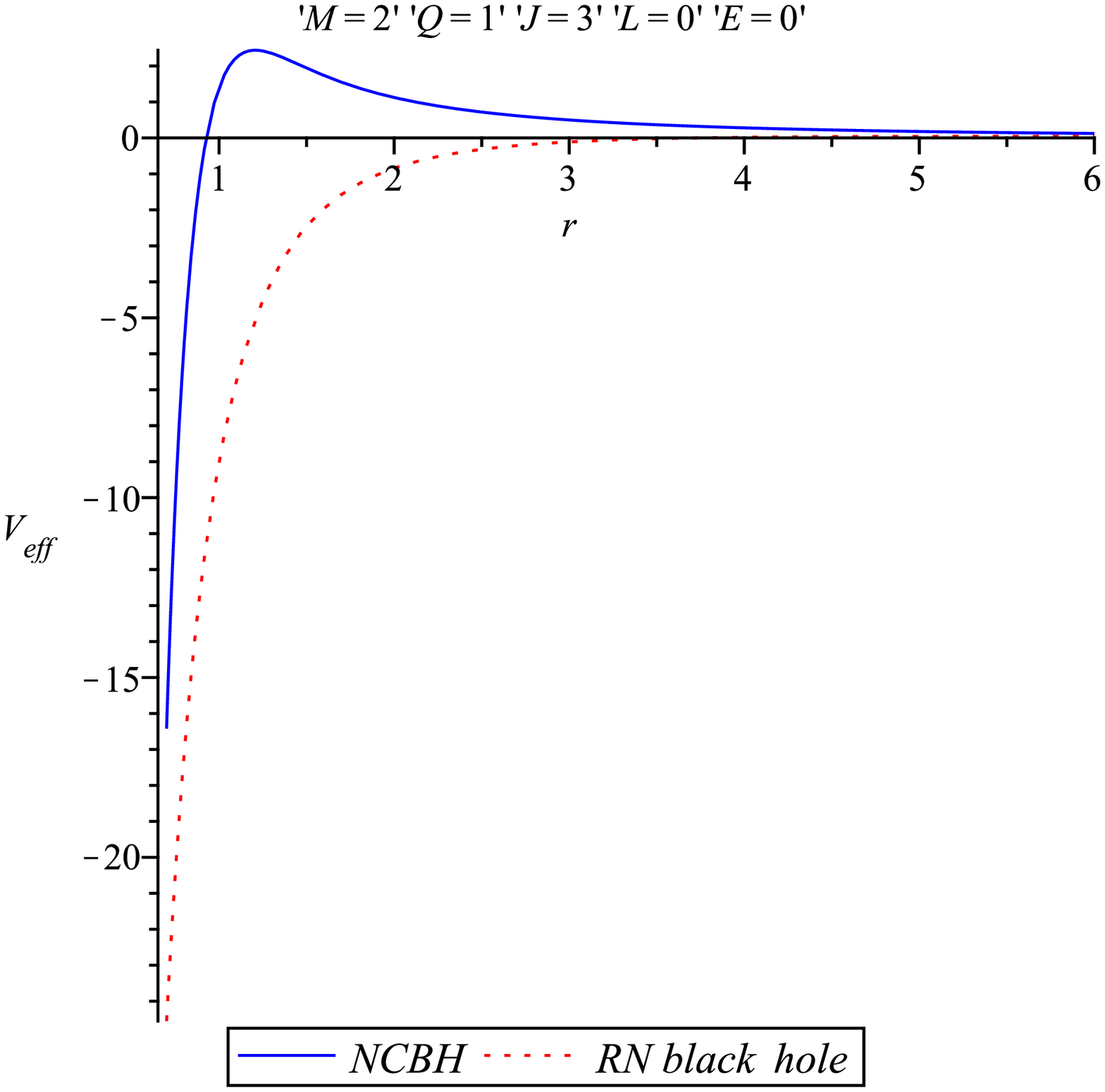}\\
\end{tabular}
\caption{ (Left) $V_{eff}$ has been plotted against $r$ for massless particle when $E \neq 0$.
 (Right) $V_{eff}$ has been plotted against $r$ for massless particle when $E =0$. }
\end{figure*}

\subsection{For Timelike particle (L=1)}

For timelike particle effective potential $V_{eff}$ becomes,
\begin{equation}
V_{eff}=-\frac{E^{2}}{2}+\frac{1}{2}\left(1+\frac{J^{2}}{r^{2}}\right)\left[1-\frac{4M}{r\sqrt{\pi}}\gamma\left(\frac{3}{2},\frac{r^{2}}{4\theta}\right)
+\frac{Q^{2}}{r^{2}\pi}\left\{\gamma^{2}\left(\frac{1}{2},\frac{r^{2}}{4\theta}\right)
-\frac{r}{\sqrt{2\theta}}\gamma\left(\frac{1}{2},\frac{r^{2}}{2\theta}\right)\right\}\right]
\end{equation}

The effective potential becomes very large as $r\rightarrow0$ and $\frac{1-E^{2}}{2}$ as  $r\rightarrow\infty$.

\begin{figure*}[thbp]
\begin{tabular}{rl}
\includegraphics[width=5.0cm]{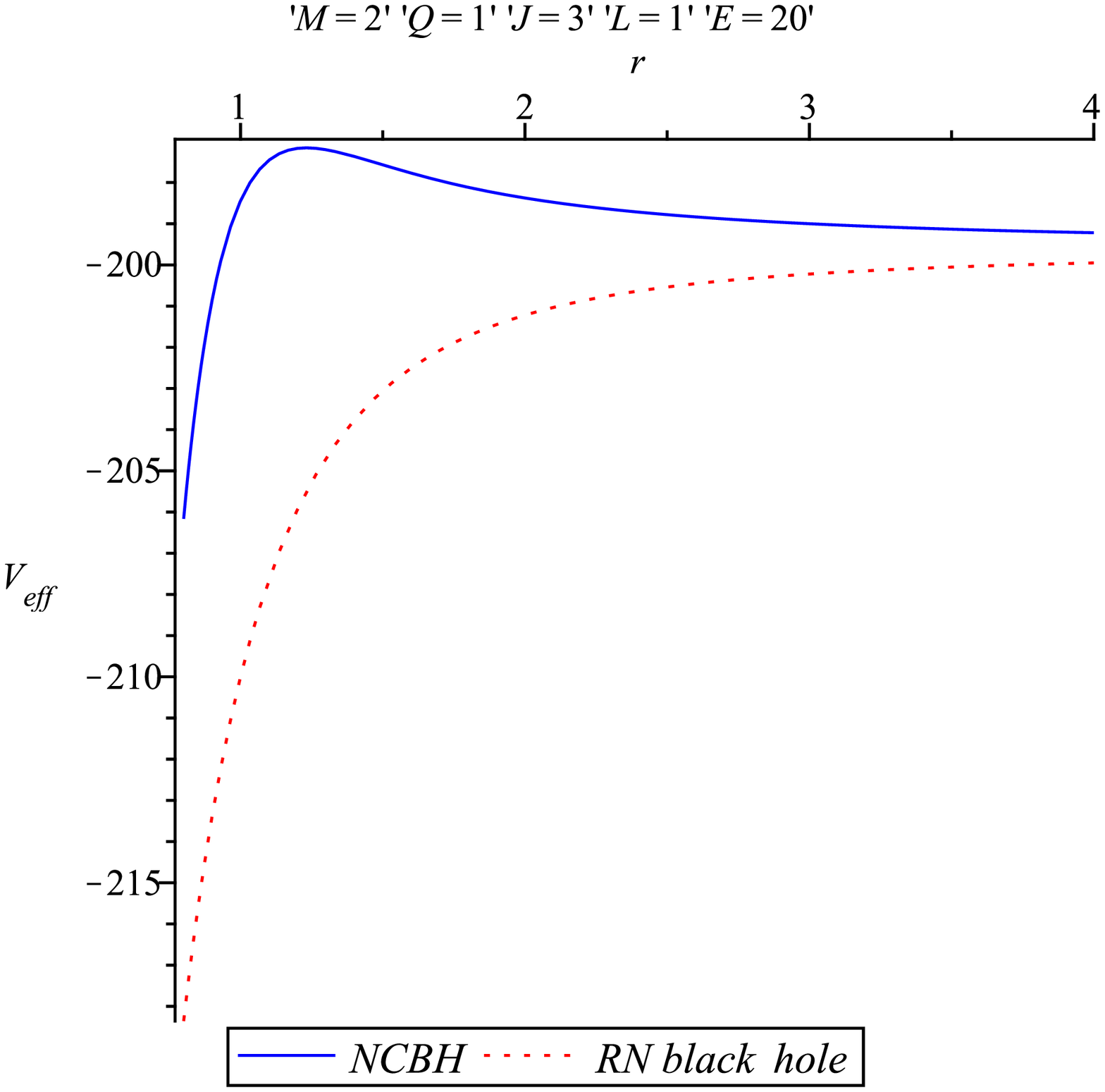}&
\includegraphics[width=5.0cm]{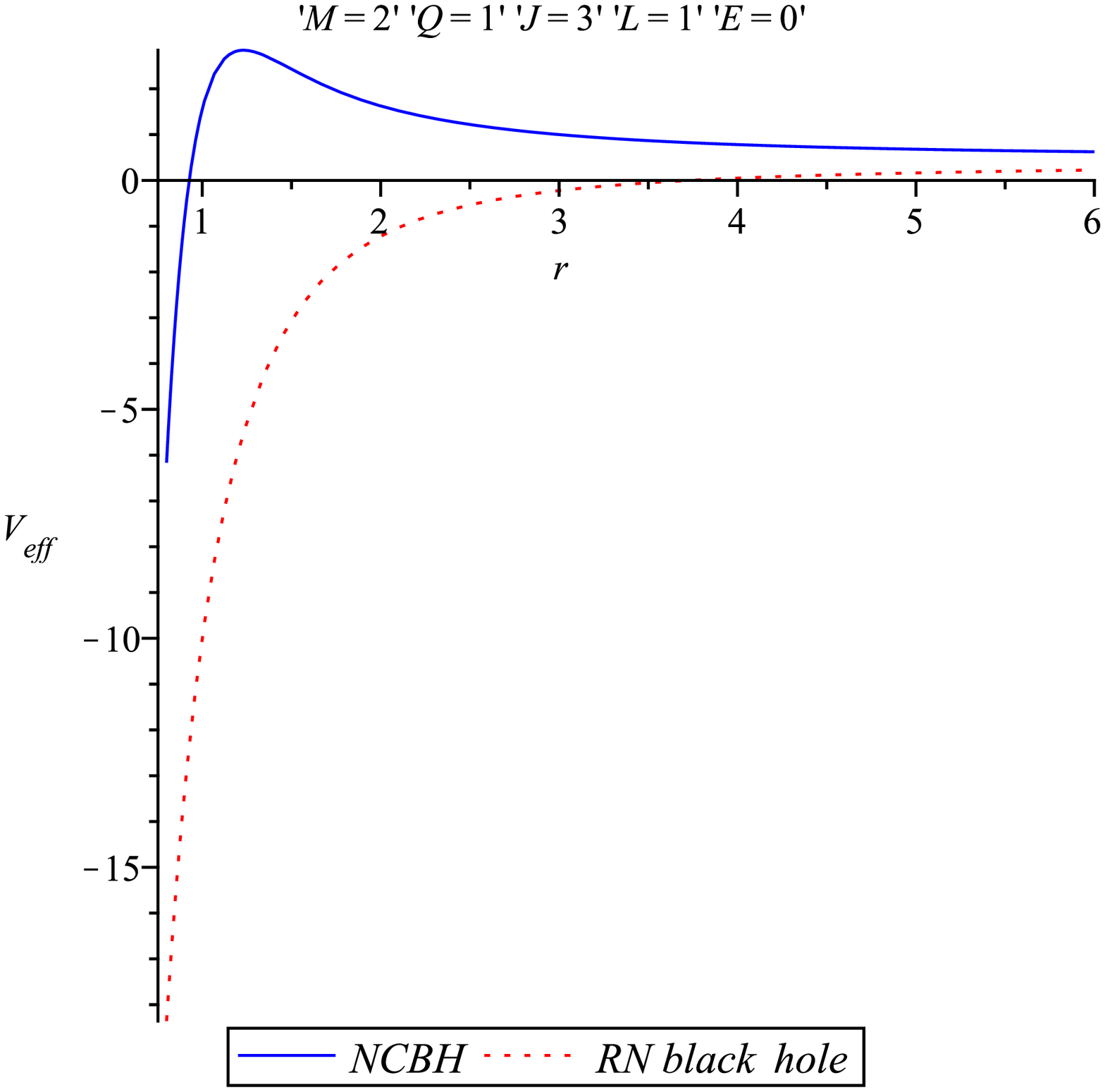}\\
\end{tabular}
\caption{ (Left) $V_{eff}$ has been plotted against $r$ for massive particle when $E \neq 0$.
 (Right) $V_{eff}$ has been plotted against $r$ for massive particle when $E =0$. }
\end{figure*}

We note that for the above choices of the parameters, say, M=2, Q=1 and $\theta=0.1$, the potential has no extremals for photon like particle, however, for massive particle, the potential has two extemals. At $r=r_h,
$
\[ V_{eff} = -\frac{E^2}{2},\]
for massless and massive particles. Also it is found that potential has a local maximum near the horizon. We should mention that for RN black hole, potential has one extremal but no local maximum point.

\section{Test Particle motion around noncommutative charged black hole}

In this section we will consider the motion of test particle around non commutative noncommutative charged black hole using Hamilton-Jacobi [ H-J ] approach. We assume that a
test particle of  mass $m$ and charge $ e $ moving in the
gravitational field of the noncommutative charged black hole. The Hamilton-Jacobi [ H-J ] equation for the test particle is given by

\begin{equation}
g^{ik} \left( \frac{\partial S}{\partial x^i} + e A_i \right) \left(
\frac{\partial S}{\partial x^k} + e A_k \right) +m^2 =0,
\end{equation}

 where  $ A_i$ is the  gauge potential and  $g_{ik}$ represents   the classical background field and S is the standard Hamilton's
characteristic function.

One can write the explicit form of H-J equation  for the metric
(2) as \cite{fr1}

\begin{equation}
-\frac{1}{f} \left( \frac{\partial S}{\partial t} + \frac{e Q}{r}
\right)^2 + f \left( \frac{\partial S}{\partial r} \right)^2 +
\frac{1}{r^2}\left( \frac{\partial S}{\partial \vartheta}\right)^2 +
\frac{1}{r^2 \sin^2 \vartheta} \left( \frac{\partial S}{\partial \phi }
\right)^2 + m^2 =0
\end{equation}

This partial equation can be solved
  using the  separation of variables method.
We   choose the $H-J$ function $ S $ in
separable form  as
\[ S(t,r,\theta,\phi) = -E.t + S_1(r) + S_2(\vartheta) + {J.\phi}, \]
 where $E$ is the energy of the particle and $J$
 is the angular momentum of the particle.

 The radial velocity of the particle takes the form
\begin{equation}
 \frac{ dr}{ dt} = f^2\left(E - \frac{eQ}{r^2}\right)^{-1}
 \sqrt{ \frac{1}{f^2} \left(E-\frac{eQ}{r} \right)^2 -
\frac{m^2}{f} - \frac{p^2}{fr^2} }
\end{equation}

The turning points of the trajectory can be obtained through the equation
$\left(\frac{dr}{dt}\right) = 0 $ and   we get

\[\left(E-\frac{eQ}{r} \right)^2 - m^2f - \frac{p^2}{r^2}f =0. \]
This equation implies
\[E= \frac{eQ}{r} + \sqrt{f} \left( m^2 + \frac{p^2}{r^2} \right)^{1/2} \]
Therefore, the    potential curve takes the form as
\[ V(r) \equiv \frac{E}{m} = \frac{eQ}{m r} + \sqrt{f} \left( 1+\frac{p^2}{m^2r^2} \right)^{1/2} \]
i.e.
\begin{equation} V(r) = \frac{eQ}{m r} +\sqrt{\left(1+\frac{p^{2}}{m^{2}r^{2}}\right)} \sqrt{1-\frac{4M}{r\sqrt{\pi}}\gamma\left(\frac{3}{2},\frac{r^{2}}{4\theta}\right)
+\frac{Q^{2}}{r^{2}\pi}\left[\gamma^{2}\left(\frac{1}{2},\frac{r^{2}}{4\theta}\right)
-\frac{r}{\sqrt{2\theta}}\gamma\left(\frac{1}{2},\frac{r^{2}}{2\theta}\right)\right]}
\end{equation}

For  a stationary situation, $ E $ i.e. $ V(r)$ should have an extremal
value. Hence  for the  extremal case
\begin{equation}
         \frac{dV}{dr} =   0,
          \end{equation}

which gives
\begin{equation}
\frac{eQ}{mr^2}\sqrt{f}\left( 1+ \frac{p^2}{m^2r^2} \right)^{1/2} =
\frac{1}{2} \left(1 + \frac{p^2}{m^2r^2} \right) f^\prime (r) -
f\frac{p^2}{m^2r^3}\end{equation}

\subsection{Test particle in Static Equilibrium (p=0)}

Note that  momentum $p$ should vanish in static equilibrium.
Therefore, we find  the value of $r$ for which potential is
extremal is given by
\begin{equation}
H(r) \equiv \frac{eQ}{mr^2}\sqrt{f}-\frac{1}{2}f'=0
\end{equation}
where $f(r)$ is given in equation $(3)$.\\
Equation $(28)$ has at least one real root where $H(r)$ cuts the $r$ axis.(see $fig-6$).
 Hence it is possible to have bound orbit for the test particle.
 In other words charged particle can be trapped by the noncommutative charged black hole in static
 equilibrium.

\begin{figure*}[thbp]
\begin{tabular}{rl}
\includegraphics[width=5.0cm]{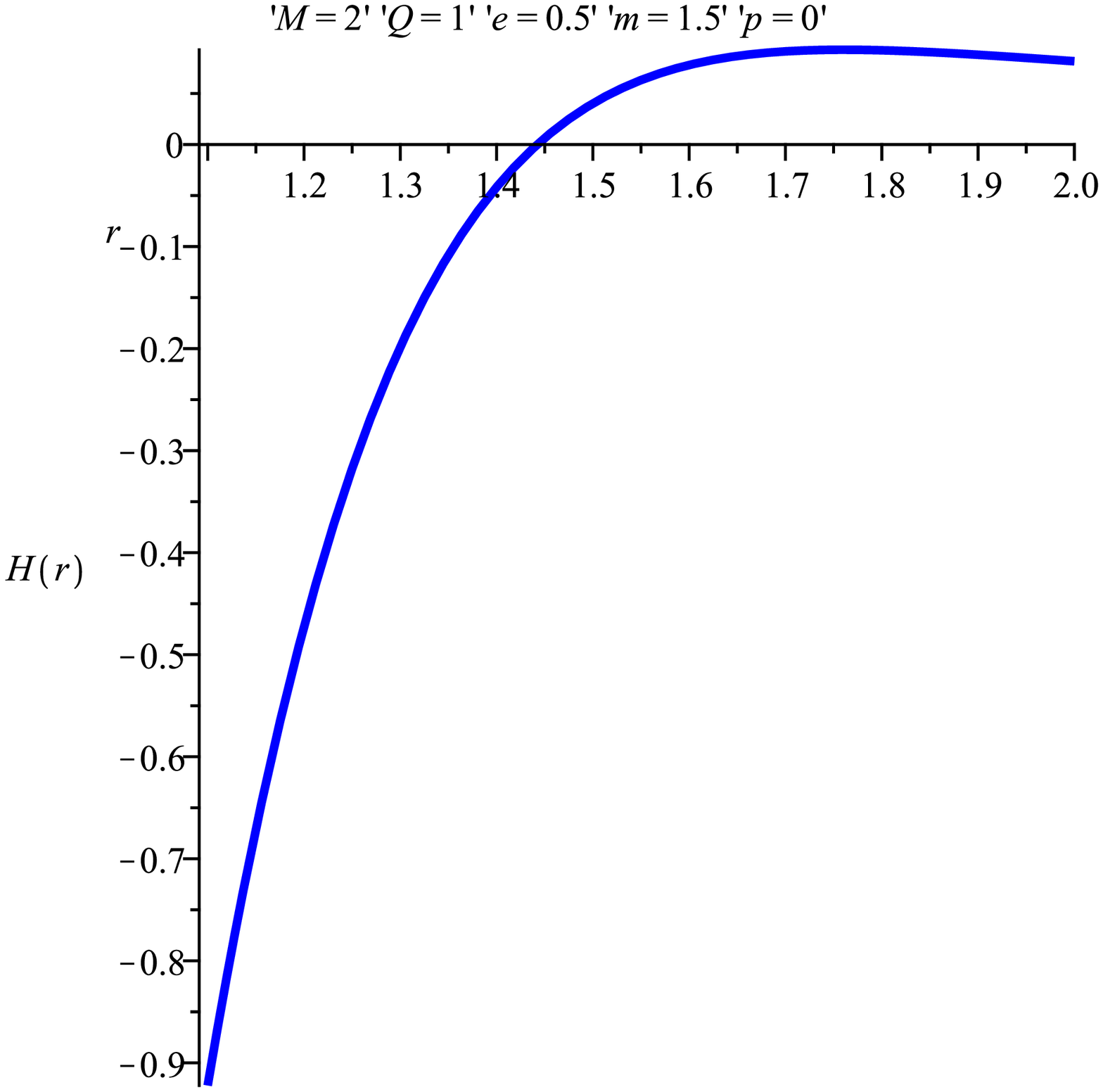}&
\includegraphics[width=5.0cm]{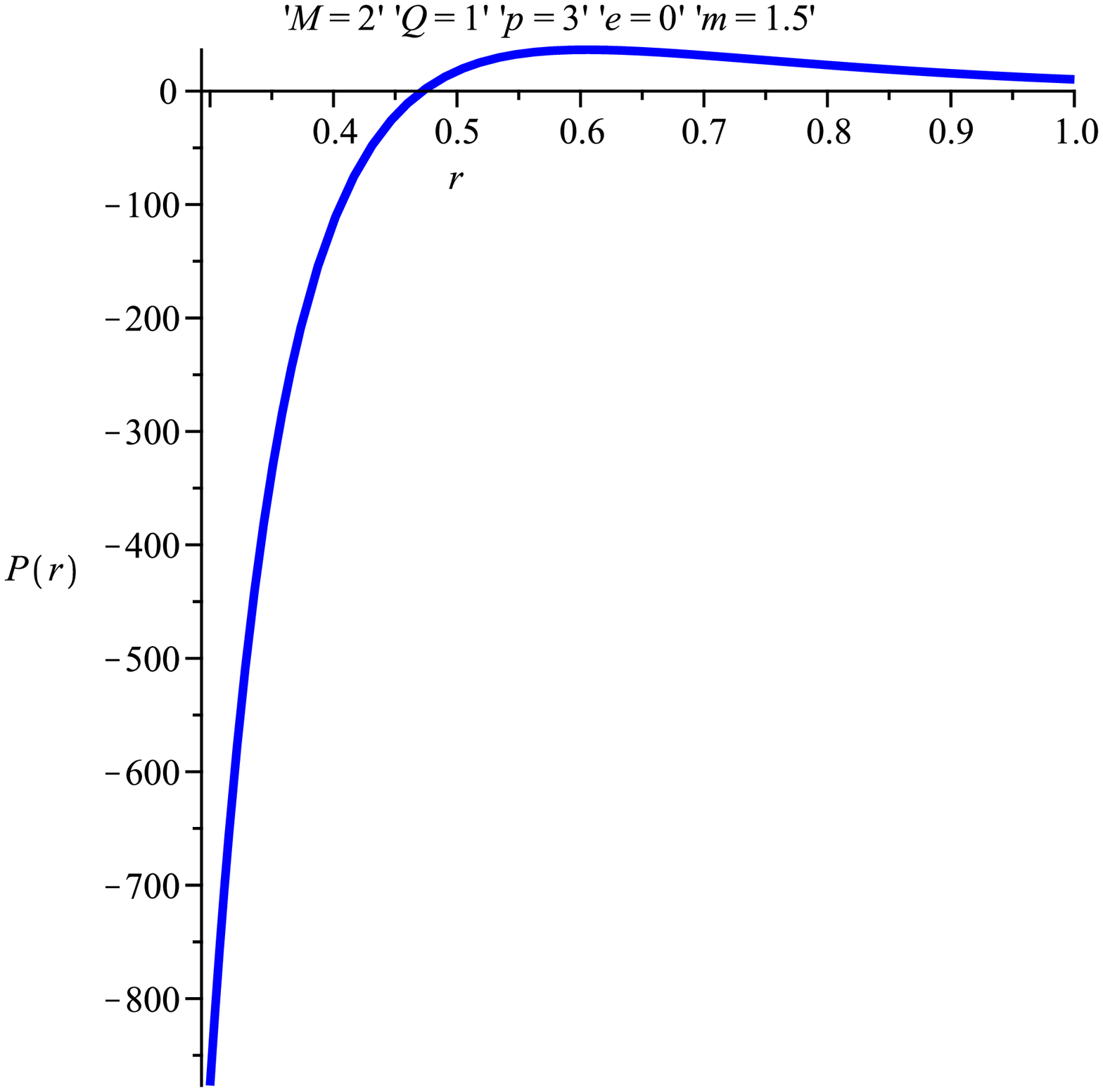}
\includegraphics[width=5.0cm]{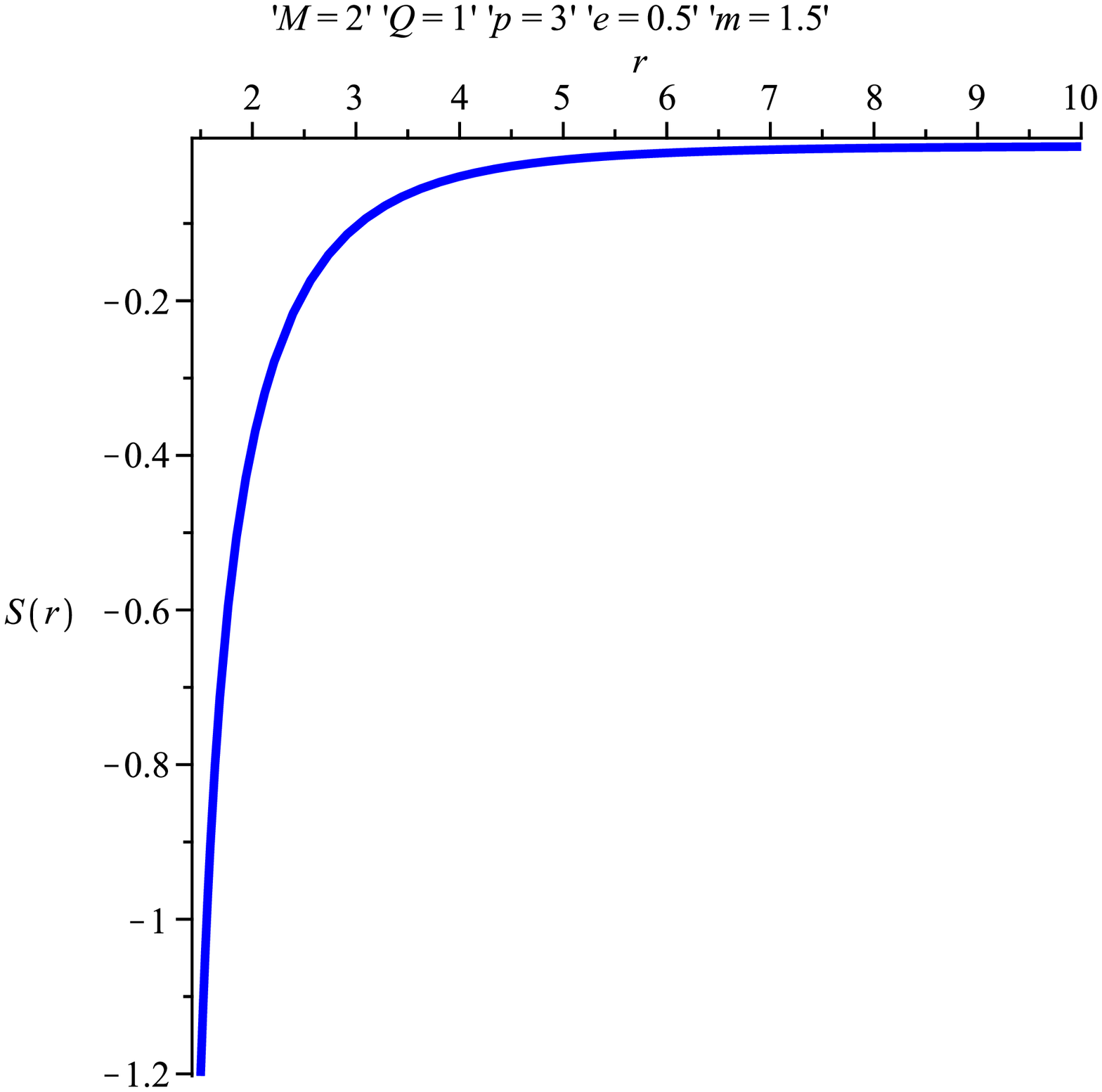}\\
\end{tabular}
\caption{ (Left) Real root exists . (Middle)   Real root exists .
 (Right) Real root does not exist. }
\end{figure*}

\subsection{Test particle in Non-Static Equilibrium }

\subsubsection{Uncharged test particle $(e=0)$}
 Now the expression $(27)$ simplifies to

\begin{equation}
P(r)\equiv \frac{1}{2}\left(1+\frac{p^{2}}{r^{2}m^{2}}\right)f'+f\frac{p^{2}}{m^{2}r^{3}}=0
\end{equation}

where $f(r)$ is given in equation $(3)$.\\

Equation $(28)$ has at least one real root where $P(r)$ cuts the
$r$ axis (see $fig-6$).  Hence it is possible to have bound orbit
for the test particle. In other words, non-charged particle can
be trapped by the noncommutative charged black hole in non-static
equilibrium.

\subsubsection{Test particle with charge $( e \neq 0 )$}

From eqn.(27), we have

\begin{equation}
S(r)\equiv\frac{eQ}{mr^2}\sqrt{f}\left( 1+ \frac{p^2}{m^2r^2} \right)^{1/2}
-\frac{1}{2} \left(1 + \frac{p^2}{m^2r^2} \right) f^\prime (r)+
f\frac{p^2}{m^2r^3}
\end{equation}

where $f(r)$ is given in equation $(3)$.\\
Equation $(29)$ has no real root since $S(r)$ does not cut the
$r$ axis (see $fig-6$). Hence it is not possible to have bound
orbit for the test particle. Hence,  charged particle can not be
trapped by the noncommutative charged  black hole in non-static
equilibrium. The possible reason for this behavior is that
noncommutative charged black hole exerts repulsive force on
charged particle.

 \section{Scattering of scalar waves in  noncommutative charged black hole geometry}\noindent
\noindent

The wave equation for   minimally coupled massless   test scalar
$\Phi $ in noncommutative charged black hole background is given
by
\begin{equation}
 \square \Phi = \frac{1}{\sqrt{-g}} \partial_\mu[\sqrt{-g}g^{\mu \nu}\partial_\nu \Phi ] =0.
\end{equation}
  Since the  noncommutative charged black hole  spacetime is spherically symmetric,
the scalar field can be separated by the variables
\begin{equation}
  \Phi_{lm}  = Y_{lm}(\theta,\phi)\frac{U_{l}(r,t)}{r}.
\end{equation}
where $Y_{lm}(\theta,\phi)$ is the spherical harmonics and $l$ is
the quantum angular momentum.\\
Using the separable form (35) in (34), we find
\begin{equation}
 \left[ \frac{1}{\sin \theta}  \frac{\partial}{\partial \theta} \sin \theta \frac{\partial}{\partial
 \theta}+
 \frac{1}{\sin^2 \theta}  \frac{\partial^2}{\partial \phi^2}\right]
 Y_{lm}= l(l+1) Y_{lm},
\end{equation}
\begin{equation}
 \ddot{U}_{l} + \frac{\partial^2U_{l}}{\partial {r^*}^2}  = V_{l} U_{l}.
\end{equation}
where the potential $V_{l}$ is given by

\[ V_{l} = \left[1-\frac{4M}{r\sqrt{\pi }}\gamma \left(\frac{3}{2},\frac{r^{2}}{4\theta
} \right)+\frac{Q^{2}}{r^{2}\pi}\left\{\gamma^{2}\left(\frac{1}{2},\frac{r^{2}}{4\theta}\right)
-\frac{r}{\sqrt{2\theta}}\gamma\left(\frac{1}{2},\frac{r^{2}}{2\theta}\right)\right\}\right]\times\]
 \[\left[  \frac{l(l+1)}{r^2} +\frac{1}{r} \left\{ \frac{4 M}{r^2 \sqrt{\pi}} \gamma
\left(\frac{3}{2},\frac{r^{2}}{4\theta }%
\right) + \frac{Mr}{\theta^{\frac{3}{2}}\sqrt{\pi}} e^{-\frac{r^2}{4
\theta}}\right\}+\frac{Q^{2}}{r^{3}\pi}
\left\{-\frac{2}{r}\gamma^{2}\left(\frac{1}{2},\frac{r^{2}}{4\theta}\right)+\frac{1}{\sqrt{2\theta}}\gamma\left(\frac{1}{2},\frac{r^{2}}{2\theta}\right)
-2\sqrt{\frac{\pi}{\theta}}e^{-\frac{r^{2}}{4\theta}}erfc\left(\frac{r}{2\sqrt{\theta}}\right)\right\}\right]\]

Here, we have  used tortoise coordinate transformation $r^*$ as
\begin{equation}
  \frac{\partial }{\partial {r^*}}  = \left[1-\frac{4M}{r\sqrt{\pi }}\gamma \left(\frac{3}{2},\frac{r^{2}}{4\theta
} \right)+\frac{Q^{2}}{r^{2}\pi}\left\{\gamma^{2}\left(\frac{1}{2},\frac{r^{2}}{4\theta}\right)
-\frac{r}{\sqrt{2\theta}}\gamma\left(\frac{1}{2},\frac{r^{2}}{2\theta}\right)\right\}\right]\frac{\partial }{\partial r}.
\end{equation}
[ dot represents the differentiation with respect to t ]
\\

The proper distance $r^*$ can be expressed as
\begin{equation}
  r^*  = \int_{r_h}^r \frac{ dx}{\left[1-\frac{4M}{x\sqrt{\pi }}\gamma \left(\frac{3}{2},\frac{x^{2}}{4\theta
} \right)+\frac{Q^{2}}{x^{2}\pi}\left\{\gamma^{2}\left(\frac{1}{2},\frac{x^{2}}{4\theta}\right)
-\frac{x}{\sqrt{2\theta}}\gamma\left(\frac{1}{2},\frac{x^{2}}{2\theta}\right)\right\}\right]} .
\end{equation}
Note that this   integration cannot be solved analytically  form,
therefore, we find the numerical values of the proper distance
$r^*$ for given values of radial distance $r$ from the horizon
radius $r_h$ ( see  Table 4 ).
\\
\begin{table}
\caption{Values of $r^*$  for different r. ($r_h = 0.9267$,
$\theta=0.1$ , $M = 2$ ,$Q=1$) } {\begin{tabular}{@{}cc@{}} \toprule $r$ &
$t$
  \\ \colrule
4 & 3.8794\\
4.5 & 4.3794  \\
5 & 4.8794 \\
5.5 & 5.3794  \\
6 & 5.8794  \\
6.5 & 6.3794  \\
7 & 6.8794 \\
 \botrule
\end{tabular}}
\end{table}

The variation of  proper distance ($r^*$) with  radial distance r  is depicted in Fig. 7.\\

\begin{figure}[htbp]
   \centering
        \includegraphics[scale=.3]{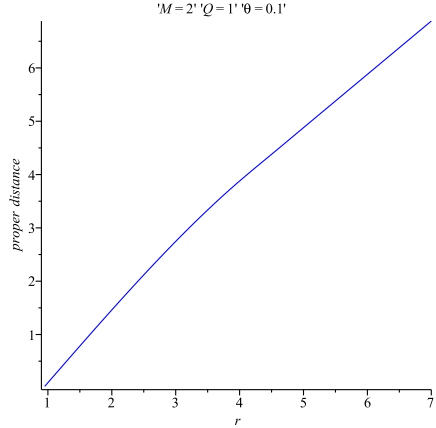}
    \caption{A plot of the  proper distance ($r^*$) versus  r.}
   \label{}
\end{figure}

For  time dependent    harmonic wave,   we can write
\begin{equation}
 U_{l} (r,t)  = \widehat{U}_{l} (r,\omega)e^{-i\omega t}
\end{equation}
Using  (38) in (35) we get    the Schr\"{o}dinger equation
\begin{equation}
\left[\frac{d^2}{d{r^*}^2} +\omega^2-  V_{l}(r) \right]
\widehat{U}_{l} (r,\omega) = 0
\end{equation}

It is observed that $V_l$ tends to zero as $ r \rightarrow \infty
$. This implies  that the solution has the type of a plane wave $
\widehat{U}_{l_0}\sim e^{\pm i \omega r^*}$ asymptotically which
is purely outgoing wave. At the horizon, $V_l \longrightarrow 0$
and is purely ingoing wave. This indicates  that if a scalar wave
is going  through the gravitational field of a  noncommutative
charged black hole, the nature of the  solution would be changed
from $e^{\pm i \omega r}$ into $e^{\pm i \omega r^*}$. In other
words,  the potential will be affected by  the scattering of
scaler waves. The potentials are depicted in figure 8.

\begin{figure}[htbp]
   \centering
        \includegraphics[scale=.4]{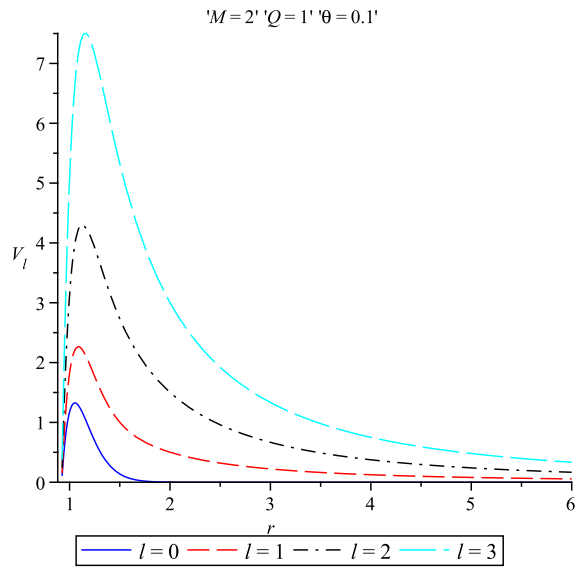}
    \caption{A plot of the  potentials of the scalar wave    for l=0,1,2,3.}
    \label{}
\end{figure}

\section{Conclusion}

Noncommutativity is coming actually  from the string theory  that
is   replaced  by the usual
 point-like structure by smeared objects. Therefore, in a natural way the  usual definition of
 Dirac-delta function   replaces everywhere with Gaussian distribution of minimal
 width $\sqrt\theta$,  where $\theta$ is  the non-commutative parameter. In our present
 article,  we have analyzed the behavior of the  time like and null geodesics of the
  non-commutative R-N black hole. The Variation of time (t) and the proper time
   ($\tau$) have been explored graphically by  plotting  against radial distance (r)
    for both massless and massive
   particle. Also, the nature of the effective potential
    has been shown by plotting against r for massless particle
    when E $\neq 0$ as well as E = 0. In case of photon like  and time like particle
     for $ E=0$ the root of the effective potential coincides with the root of the
     non-commutative RN black hole. From our analysis it is shown    that for both case of
      charged particle in static equilibrium and uncharged test particle in non-Static
       equilibrium ,
       particles can be trapped by the noncommutative charged black hole , where as
       in case of uncharged particle in non static equilibrium it is not possible
       to have bound orbit for the test particle. So uncharged particle can not be
        trapped by the noncommutative charged black hole in the case of  non-static
         equilibrium. The possible reason for this behavior  is that noncommutative
         charged black hole exerts repulsive force on uncharged particle.  We have also
          compared the behavior of the time-like and null geodesics between Non-commutative
           R-N  Black Hole and R-N black hole.  It is also  shown that for
            $r\rightarrow \infty$ the   non-commutative R-N  Black
            hole comes back to the R-N black hole. Finally we studied the
             scattering problem of the scalar wave in  noncommutative
              R-N black hole geometry. It is seen
               that the potential will be affected by  the scattering of
scaler waves. The characteristic of the potential is shown
graphically.
                              For $Q=0$ our  results coincide  with non charged case \cite{fr}.  \\

\begin{center}
  {\Large  \textsc{ Acknowledgments} }
\end{center}

          FR and RB  would   like to thank   Inter-University Centre for Astronomy and Astrophysics (IUCAA),
          Pune, India for providing research facilities.       \\

\end{document}